\RequirePackage[OT1]{fontenc}
\documentclass[10pt,conference,final]{IEEEtran}
\IEEEoverridecommandlockouts
\usepackage{cite}
\usepackage{amsmath,amssymb,amsfonts}
\usepackage{graphicx}
\usepackage{textcomp}
\usepackage{xcolor}
\usepackage{subcaption}

\usepackage{algorithmicx}
\usepackage{algorithm}
\usepackage[noend]{algpseudocode}
\algnewcommand\algorithmicfunctionname{\textbf{function:}}
\algnewcommand\Functionname{\item[\algorithmicfunctionname]}
\algnewcommand\algorithmicinput{\textbf{Input:}}
\algnewcommand\Input{\item[\algorithmicinput]}
\algnewcommand\algorithmicoutput{\textbf{Output:}}
\algnewcommand\Output{\item[\algorithmicoutput]}
\newcommand{\algorithmicbreak}{\textbf{break}}
\newcommand{\Break}{\algorithmicbreak}
\newcommand{\algorithmiccontinue}{\textbf{continue}}
\newcommand{\Continue}{\algorithmiccontinue}

\DeclareMathOperator*{\argmin}{argmin} 

\usepackage[english]{babel}
\usepackage{blindtext}
\usepackage{units}

\usepackage{booktabs}
\usepackage{multirow}
\newcommand{\ra}[1]{\renewcommand{\arraystretch}{#1}}
\usepackage{diagbox}

\def\BibTeX{{\rm B\kern-.05em{\sc i\kern-.025em b}\kern-.08em
    T\kern-.1667em\lower.7ex\hbox{E}\kern-.125emX}}

\usepackage[bookmarks=false]{hyperref}

\begin{document}
\bstctlcite{IEEEexample:BSTcontrol}

\title{Energy-efficient Runtime Resource Management for Adaptable Multi-application Mapping}

\author{\IEEEauthorblockN{Robert Khasanov\IEEEauthorrefmark{1}, Jeronimo Castrillon\IEEEauthorrefmark{2}}
\IEEEauthorblockA{{Chair for Compiler Construction}, {Technische Universit\"at Dresden}, Germany \\
Email: \IEEEauthorrefmark{1}robert.khasanov@tu-dresden.de, \IEEEauthorrefmark{2}jeronimo.castrillon@tu-dresden.de}
}

\maketitle

\begin{abstract}
  Modern embedded computing platforms consist of a high amount of heterogeneous resources, which allows executing multiple applications on a single device.
The number of running application on the system varies with time and so does the amount of available resources.
This has considerably increased the complexity of analysis and optimization algorithms for runtime mapping of firm real-time applications. 
To reduce the runtime overhead, researchers have proposed to pre-compute partial mappings at compile time and have the runtime efficiently compute the final mapping. 
However, most existing solutions only compute a fixed mapping for a given set of running applications, and the mapping is defined for the entire duration of the workload execution.
In this work we allow applications to adapt to the amount of available resources by using mapping segments.
This way, applications may switch between different configurations with varied degree of parallelism.
We present a runtime manager for firm real-time applications that generates such mapping segments based on partial solutions and aims at minimizing the overall energy consumption without deadline violations.
The proposed algorithm outperforms the state-of-the-art approaches on the overall energy consumption by up to $13\%$ while incurring an order of magnitude less scheduling overhead.

\end{abstract}


\section{Introduction}
\label{sec:intro}
Most modern computing systems are embedded in end-user devices.
Some of these systems consist of many-cores, and the number of cores have already reached the thousands~\cite{olofsson2016epiphany}.
Many-cores allow executing multiple applications in parallel on a single device.
These applications are launched at any time, making resource management particularly challenging.

The problem of optimally executing multiple applications is well-known in the embedded domain.
Existing solutions might be classified into design-time, runtime and hybrid~\cite{Singh2013}.
The latter combines the benefits of the first two approaches.
In a hybrid mapping, the decision where to execute an application is split between design- and runtime.
At design time, partial solutions for each application are generated by analyzing the applications in isolation.
This is done, e.g., using established methodologies for Design Space Exploration (DSE).
At runtime, the runtime manager (RM), 
being aware of the overall workload, transforms these solutions into final mappings.
Hybrid mappings thus benefit from extensive design space exploration to find near-optimal partial solutions, and can adapt to the workload at runtime via efficient heuristics.  

In runtime and hybrid approaches, when a new application arrives, the RM has to assign resources to it.
An incremental RM allocates the new application on free resources~\cite{Singh13,daarm}.
If available resources do not suffice, the RM either rejects the application, or uses fast heuristics to remap existing jobs.
For instance in~\cite{Ykman06,wildermann15}, a joint mapping is computed for all applications at once.
Authors formulate the problem as multi-choice multidimensional knapsack problem (MMKP)~\cite{knapsack_book} and solve it using fast heuristics.
Similarly, when an application finishes execution, more resources become available and the RM can generate new mappings.
State-of-the-art solutions, as discussed above, generate \emph{fixed} mappings, i.e., mappings that do not change during the execution of the applications.
Those fixed mappings are optimized \textit{locally} for the duration of the fixed set of applications, and at the subsequent RM activations, new local optimal mappings will be generated.
However, a sequence of local optimal decisions might lead to a sub-optimal schedule in the scope of the entire runtime of the system.

Recently, Niknafs et al.~\cite{niknafs19} attempted to enlarge the scope of analysis for firm real-time applications by computing joint mappings. 
In the generated schedules, different applications might be mapped to the same core and executed according to the earliest deadline first (EDF) policy.
Applications might be also preempted in favour of more critical and proactively predicted jobs.
However, the presented approach is limited to single-threaded applications.
In this paper we generalize Niknafs' approach to multi-threaded applications.
We express the schedules as fragmented into mapping segments, which explicitly express resource adaptations (Section~\ref{sec:problem}).
We propose a fast algorithm for firm real-time multi-threaded applications, which analyzes the applications in the scope until the last application finishes,
   and generates the mapping segments optimized w.r.t. the overall energy consumption (Section~\ref{sec:algo}).
Due to the enlarged scope of the analysis, the generated schedules will be near-optimal for the entire execution of the current workload (Section~\ref{sec:eval}).

\section{Related work}
\label{sec:relwork}
The problem of optimal execution of multiple applications is well-known. 
The common aim is to increase the overall throughput by using historical data~\cite{Gregg11, castrill_multikpn2010},
or by exploiting the concavity of throughput of multi-threaded applications~\cite{Venkataramani19}.
These works do not optimize energy efficiency.

Several runtime managers for energy efficient execution of applications exist.
Das et al.~\cite{Das16} uses energy-awareness of single applications to improve both energy consumption and thermal dissipation.
Greedy heuristics are used for thread allocations, and reinforcement learning is employed for selecting the minimum frequency.
Tzilis et al.~\cite{Tzilis19} uses profiling data to predict the performance and energy consumption of a single-threaded application co-scheduled with another applications, and decides on application placement and frequency settings. 
Similarly, Libutti et al.~\cite{Libutti16} exploits collected off-line information such as CPU demands and memory sensitivity for job co-scheduling.
Libutti et al. target multi-threaded applications but without controlling frequency settings.
Singh et al.~\cite{Singh16} capture the trace information of individual applications at design-time, and then merge execution intervals of multiple applications at runtime.
This approach is limited to Synchronous Data Flow Graphs (SDFG) and thus cannot be applied to dynamic workloads.

Unlike aforementioned works, hybrid approaches prepare a set of partial or complete mappings of individual applications at design-time, which correspond to Pareto-optimal configurations.
Ascia et al.~\cite{Ascia04} were one of the first to propose multi-objective mapping generation,
using evolutionary algorithms to find Pareto-optimal points.
Approaches such as~\cite{Massari14,Singh13,quan2015hybrid,Onnebrink19} use heuristics for efficient exploration of Pareto-optimal configurations, while others such as \cite{Mariani12,daarm} use evolutionary algorithms.
The amount of considered configuration can be also reduced by exploiting system symmetries~\cite{scopes17_tetris,goens_taco17}.

Identified Pareto-optimal points at design time are passed to the RM.
The essential task of the RM is to select the operating points for multiple applications.
Singh et al.~\cite{Singh13} iteratively map applications onto the platform, whereas Weischlgartner et al.~\cite{daarm} enhance the iterative binding of application with a repair heuristic.
More sophisticated approaches express the problem as a multiple-choice multidimensional knapsack problem (MMKP)~\cite{knapsack_book}.
Ykman-Couvreur et al.~\cite{Ykman06}, for instance, propose a fast heuristic, which expresses the resource demands of operating points as a single value, and then use a greedy algorithm to solve the MMKP. 
This heuristic underlies the solutions proposed in~\cite{Massari14,Mariani12}.
Wildermann et al.~\cite{Wildermann14,wildermann15} solve the problem by applying a Lagrangian relaxation method.
Shojaei et al.~\cite{Shojaei13} propose a compositional Pareto-algebraic heuristic using Pareto-algebra.
However, these algorithms assume that applications are constantly running, and do not consider application reconfiguration.
As mentioned in Section~\ref{sec:intro}, Niknafs et al.~\cite{niknafs19} do consider application reconfiguration, but limited to single-threaded applications.

\section{Motivational example}
\label{sec:motivation}
\begin{table}
\centering
\ra{1.1}
\caption{Request parameters.}
\label{tab:motiv_reqs}
\begin{tabular}{@{}cccccccc@{}}
\toprule
	Req. & App. & \phantom{abc} & S1-Arr. & S1-Dead. & \phantom{abc} & S2-Arr. & S2-Dead. \\
\midrule
	$\sigma_1$ & $\lambda_1$ && 0 & 9 && 0 & 9 \\
	$\sigma_2$ & $\lambda_2$ && 1 & 5 && 1 & 4 \\
\bottomrule
\end{tabular}
\vspace{-3mm}
\end{table}

\begin{table}
\centering
\ra{1.1}
\caption{Application parameters.}
\label{tab:motiv_apps}
\begin{tabular}{@{}ccccccc@{}}
\toprule
 &&  \multicolumn{2}{c}{$\lambda_1$, pr. 0\% - 18.87\% - 62.08\% } & \phantom{a} & \multicolumn{2}{c}{$\lambda_2$, pr. 0\%} \\
   \cmidrule{3-4} \cmidrule{6-7}
\#L & \#B & $\tau [s]$  & $\xi [J]$ && $\tau [s]$  & $\xi [J]$ \\
\midrule
	1 & 0 & 16.8 - 13.63 - 6.37 & 7.90 - 6.41 - 3.00   && 10.0 & 2.00 \\
	2 & 0 & 10.3 - 8.36 - 3.91  & 7.01 - 5.69 - 2.66   && 7.0 & 2.87 \\
	0 & 1 & 11.2 - 9.09 - 4.25  & 18.54 - 15.04 - 7.03 && 5.0 & 7.55 \\
	0 & 2 & 6.3 - 5.11 - 2.39   & 17.70 - 14.36 - 6.71 && 3.5 & 10.5 \\
	1 & 1 & 8.1 - 6.57 - 3.07   & 10.90 - 8.84 - 4.13  && 3.5 & 6.44 \\
	1 & 2 & 7.9 - 6.41 - 3.00   & 10.60 - 8.60 - 4.02  && 3.0 & 6.81 \\
  2 & 1 & 5.3 - 4.30 - 2.01   & \underline{8.90} - 7.22 - 3.38   && 3.0 & 5.73 \\
	2 & 2 & 4.7 - 3.81 - 1.78   & 11.00 - 8.92 - 4.17  && 2.0 & 6.58 \\
\bottomrule
\end{tabular}
\vspace{-3mm}
\end{table}

Assume a heterogeneous multi-core device with 2 little and 2 big cores that serves requests arriving according to Scenario~S1 in Table~\ref{tab:motiv_reqs}.
Each request consists of the application to run, its arrival and (absolute) deadline.
Table~\ref{tab:motiv_apps} describes configurations of the two applications of the example ($\lambda_1, \lambda_2$), characterized by the number of little and big cores, the execution time $\tau$ and the energy consumption $\xi$.
For $\lambda_1$ we show time/energy values in triples, in which the first (initial) state is followed by the states states with a progress ratio of $18.87\%$ and $62.08\%$ correspondingly,
to which we refer below.
The values in the table are synthetic, but feature ratios similar to what we observed in real applications (see Section~\ref{sec:eval}).

\begin{figure}[b]
  \vspace{-5mm}
  \centering
  \includegraphics[width=0.5\textwidth]{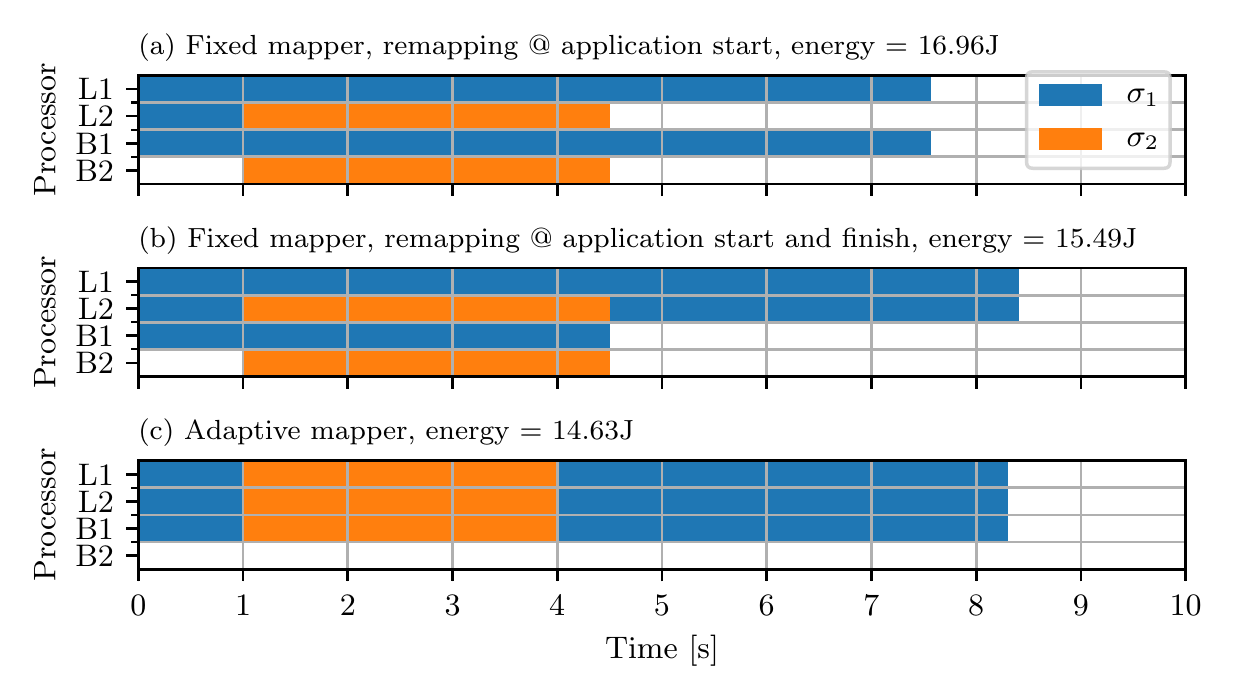}
  \vspace{-8mm}
  \caption{Three resource management scenarios.}
  \vspace{-2mm}
  \label{fig:motiv_scn}
\end{figure}

At $t = 0$, the RM receives request $\sigma_1$ to execute application $\lambda_1$.
An energy-optimizing mapper decides to map it to 2 little and 1 big cores (i.e., \texttt{2L1B}),
since this configuration meets the deadline ($t=9$) with the least energy consuming ($\unit[8.9]{J}$, underlined).
After \unit[1]{s}, request $\sigma_2$ arrives while request $\sigma_1$ progressed to $\approx 18.87\%$. 
To meet the deadline, $\sigma_2$ must be executed either as \texttt{2B}, \texttt{1L1B}, \texttt{1L2B}, \texttt{2L1B} or \texttt{2L2B}.
If any of these mappings is chosen, $\sigma_1$ must then continue as \texttt{1L}, \texttt{2L}, \texttt{1L1B} or \texttt{1B}.
A mapper that only explores fixed mappings would choose one where both jobs meet their deadlines, e.g., \texttt{1L1B} for both $\sigma_1$ and $\sigma_2$.
By $t=4.5$, $\sigma_2$ finishes its execution and $\sigma_1$ progresses to $\approx 62.08\%$.
If $\sigma_1$ continues as \texttt{1L1B} till the end, as depicted in Fig.~\ref{fig:motiv_scn}(a), the overall energy consumption is \unit[16.96]{J}.
If the RM decides to remap application at $t=4.5$, then it would choose the most efficient mapping \texttt{2L}, see Fig.~\ref{fig:motiv_scn}(b), with overall energy consumption of \unit[15.49]{J}.
However, if at $t=1$ the RM runs $\sigma_2$ on \texttt{2L1B} and suspends momentarily $\sigma_1$, then when $\sigma_2$ finishes, $\sigma_1$ may continue with \texttt{2L1B} leading to an overall energy consumption of \unit[14.63]{J} (see Fig.~\ref{fig:motiv_scn}(c)).

Assume now the tighter Scenario~S2 in Table~\ref{tab:motiv_reqs}.
At $t=1$, $\sigma_2$ can only choose \texttt{1L2B}, \texttt{2L1B}, or \texttt{2L2B}, which leaves at most either 1 big or 1 little core for $\sigma_1$.
Since these configurations cannot meet the deadline, a fixed mapper will be unable to find a schedule and  $\sigma_2$ will be rejected.
With explicit adaptations, a dynamic mapper in the RM can produce the schedule in Fig.~\ref{fig:motiv_scn}(c) and meet the constraints.
By allowing mapping reconfigurations and global analysis scope, the overall energy consumption and the request acceptance rate can be improved.

\section{System model and problem definition}
\label{sec:problem}
To formalize the resource management problem illustrated above, assume a heterogeneous platform with $m$ resource types, and core counts represented by the vector $\vec{\Theta} = (\Theta_1, \dots, \Theta_{m})$.
The platform executes multi-threaded applications, in which each thread perform computations during the whole execution of the application.
Additionally, we assume that in each fixed configuration all threads process the workload with a constant progress rate.
Via a DSE method or benchmarking, the RM is given information about each application $\lambda$ and its $N_\lambda$ operating points (cf. Table~\ref{tab:motiv_apps}).
Each operating point consists of needed resources $\vec{\theta}$, the (worst case) execution time $\tau$ and the energy consumption $\xi$, i.e., $c_\lambda^j = \langle \vec{\theta}, \tau, \xi \rangle$.
Operating points are assumed to be already Pareto-filtered, i.e., each operating point is better than any other in at least one parameter.

Every time a request arrives, the RM is activated. 
We denote by $\Sigma_{t'}$
the set of unfinished jobs admitted in $t < t'$ plus the newly arrived job at time $t'$.
For a given job, $\sigma[\alpha]$ denotes the arrival time, 
$\sigma[\delta]$ the (absolute) deadline, 
$\sigma[\lambda]$ the application, and 
$\sigma[\rho] \in [0,1]$ the remaining progress ratio of the job, i.e., $\sigma = \langle \alpha, \delta, \lambda, \rho \rangle$.
Given a set of requests $\Sigma_{t'}$, the RM attempts to find a schedule $\kappa$, which is defined as a list of mappings $\mu$ defined on consecutive time segments: 
\begin{equation}
\kappa =  \{\mu_i \times \Delta_{\mu_i}\}, \qquad 0 \leq i < N\label{eq:sched_def}\\
\end{equation}
  where $N$ is a number of segments, ${\Delta_\mu  =  [\underline{\Delta_\mu}, \overline{\Delta_\mu})}$ is a time interval of the mapping $\mu$, ${\underline{\Delta_{\mu_{0}}} = t'}$ and ${\overline{\Delta_{\mu_i}} = \underline{\Delta_{\mu_{i+1}}}}$, ${0 \leq i < N-1}$.
Each mapping $\mu$ contains individual job mappings $\nu = \langle \sigma, \lambda, j \rangle$, which expresses that application $\lambda$ of request $\sigma$ runs with configuration $j$. 
Note that this formulation allows jobs to be remapped from one configuration to another in the schedule.

Having introduced the notation, the optimization problem can be defined as:

\begin{subequations}
\begin{align}
\text{minimize} & \sum_{\mu \in \kappa} \sum_{\nu \in \mu} c_{\nu[\lambda]}^{\nu[j]}[\xi] \frac{\left| \Delta_\mu \right|}{c_{\nu[\lambda]}^{\nu[j]}[\tau]} \\
  \text{subject to} & \sum_{\nu \in \mu} c_{\nu[\lambda]}^{\nu[j]}[\vec{\theta}] \leq \vec{\Theta},  \qquad \forall \mu \in \kappa \label{eq:resource}\\
 & \nu_1[\sigma] \neq \nu_2[\sigma],\qquad \forall \nu_1,\nu_2 \in \mu,  \nu_1 \neq \nu_2 \label{eq:non-twise}\\ 
 & \sum_{\substack{\mu \in \kappa \\ \nu \in \mu \\ \nu[\sigma] = \sigma}}
        \frac{\left| \Delta_\mu \right|}{c_{\nu[\lambda]}^{\nu[j]}[\tau]} = \sigma[{\rho}], \qquad \forall \sigma \in \Sigma \label{eq:full_time}\\
 & \max_{\substack{\mu \in \kappa\\ \nu \in \mu\\ \nu[\sigma] = \sigma}}
        \overline{\Delta_\mu} \leq \sigma[\delta], \qquad \forall \sigma \in \Sigma. \label{eq:deadline}
\end{align}
\end{subequations}
Constraint~\eqref{eq:resource} ensures that the resources required for mapping in the segment $\mu$ does not exceed the available resources per type $\vec{\Theta}$.
Equation~\eqref{eq:non-twise} specifies that at each mapping segment at most one job mapping relates to each request $\sigma$.
Constraint~\eqref{eq:full_time} expresses that each job will run until the end.
Finally, each job must finish the execution before the deadline~\eqref{eq:deadline}.
If there is a feasible solution to this problem, the RM admits the request and changes the schedule accordingly. Otherwise the request is rejected.

\section{Fast heuristic for MMKP-based scheduling}
\label{sec:algo}
\begin{algorithm}
  \caption{MMKP-MDF mapping heuristic.}
  \label{algo:mmkp-mdf}
  \begin{algorithmic}[1]
    \Input{Set of jobs $\Sigma_{t}$, platform $\vec{\Theta}$, application table $c$}
    \Output{The schedule $\kappa$}
    \State $\vec{J} \gets \vec{\Theta} \times \operatorname{max}({\sigma[\delta] - t \mid \sigma \in \Sigma_t}) $ \label{algo:mdf:init_cont}
    \ForAll {$\sigma \in \Sigma_{t}$} $jc[\sigma]  \gets \varnothing$ \label{algo:mdf:init_jc}
    \EndFor
    \While{$\exists \sigma \in \Sigma_t: jc[\sigma] = \varnothing $} \label{algo:mdf:main_loop}
      \State $\sigma^*, cl \gets $ \Call{NextJobMDF}{$\Sigma_t, jc, \vec{J}, c$} \label{algo:mdf:next_job}
      \While {$jc[\sigma^*] = \varnothing$} \label{algo:mdf:map_job_begin}
        \If {$\left| cl \right| = 0$ }
          \Return $\varnothing$ \label{algo:mdf:no_feas_mapping}
        \EndIf
        \State $ j^* \gets \argmin_{j \in cl} \{ c_{\sigma^*[\lambda]}^j[\xi] \} $
        \State $jc^* \gets jc$, $jc^*[\sigma^*] \gets j^* $
        \State $\kappa^* \gets$ \Call{ScheduleJobs}{$\Sigma_{t}, jc^*, \vec{\Theta}, c$} \label{algo:mdf:schedule_jc}
        \If {$\kappa^* \neq \varnothing$}
          \State $jc \gets jc^*$, 
                 $\kappa \gets \kappa^*$ \label{algo:mdf:scheduled_job_begin}
          \State $\vec{J} \gets \vec{J} - c_{\sigma^*[\lambda]}^{j^*}[\vec{\theta}] \times c_{\sigma^*[\lambda]}^{j^*}[\tau] \times  \sigma^*[\rho] $
          \label{algo:mdf:scheduled_job_end}
        \Else
          \State $cl \gets cl \setminus j^*$ 
        \EndIf
      \EndWhile \label{algo:mdf:map_job_end}
    \EndWhile
    \State \Return $\kappa$
  \end{algorithmic}
\end{algorithm}

We consider the core types as knapsacks with certain \textit{capacities},
and the job configurations $c_\lambda^j$ as \textit{items} with certain \textit{weights}.
Weights are defined as the required processing time to finish the job $c_\lambda^j[\tau]$ times the required number of resources of corresponding type $c_\lambda^j[\vec{\theta}]$, while the capacities express the available processing time per each resource type.
Each job forms a group of items, in which exactly one item must be chosen.
After negating the energy \textit{values} of each item, the optimization goal becomes to maximize the overall (negative) value which can be expressed as a multiple-choice multi-dimensional knapsack problem (MMKP)~\cite{knapsack_book}.
Algorithm~\ref{algo:mmkp-mdf} describes our heuristic to solve this.
It generalizes the solution in~\cite{niknafs19} for multi-threaded applications.

At each activation of the RM, the containers $\vec{J}$ are initialized with the overall processing time per resource type limited by the largest job deadline (the time scope of the analysis), which is followed by the initialization of the found job configuration dictionary $jc$ (lines~\ref{algo:mdf:init_cont}-\ref{algo:mdf:init_jc}). 
The algorithm iterates over unmapped jobs (line~\ref{algo:mdf:main_loop}), using \textproc{NextJobMDF} to select the next one to map (line~\ref{algo:mdf:next_job}).
Similarly to~\cite{niknafs19} this function 
(i) filters configurations by checking whether they can meet deadlines and fit the containers $\vec{J}$, 
(ii) determines a job, in which the difference between the most energy-efficient feasible configuration and the second best one is maximized, i.e., Maximum Difference First (MDF), and 
(iii) returns the selected job $\sigma^*$ along with a list of filtered configurations $cl$.
The MDF policy prioritizes the job that would cause the highest degradation if the best point is not chosen in this iteration.
In lines~\ref{algo:mdf:map_job_begin}-\ref{algo:mdf:map_job_end}, the algorithm iterates 
over the configurations in non-decreasing order of energy consumption.
It then attempts to schedule the job $\sigma^*$ with already mapped jobs in \textproc{ScheduleJobs},
detailed in Algorithm~\ref{algo:schedule_jc}.
Once the job is successfully scheduled, its configuration and a new schedule is saved, and the containers $\vec{J}$ are updated (lines~\ref{algo:mdf:scheduled_job_begin}-\ref{algo:mdf:scheduled_job_end}).
Otherwise, no feasible schedule was found and the algorithm exits in line~\ref{algo:mdf:no_feas_mapping}.

\begin{algorithm}
  \caption{Schedule jobs.}
  \label{algo:schedule_jc}
  \begin{algorithmic}[1]
    \Input{Set of jobs $\Sigma_t$, their configurations $jc$, platform $\vec{\Theta}$, application table c }
    \Output{The schedule $\kappa$}
    \State $\widetilde{\Sigma} \gets \{ \sigma \in \Sigma_t \mid jc[\sigma] \neq \varnothing \} $,
           $\kappa \gets \varnothing$, $t^e \gets t$ \label{algo:sj:init}
    \While {$ \left| \widetilde{\Sigma} \right| \neq 0 $} \label{algo:sj:main_loop_begin}
      \State $\sigma^* \gets \argmin_{\sigma \in \widetilde{\Sigma} } \{ \sigma[\delta]  \} $ \label{algo:sj:next_job_edf}
      \State $j^* \gets jc[\sigma^*]$,
            $\rho^* \gets \sigma^*[\rho]$ \label{algo:sj:rr_init}

      \ForAll { $\mu \times \Delta \in \kappa $} \label{algo:sj:segments_loop_begin}
        \State $\vec{\theta^*} \gets \sum_{\nu \in \mu} c_{\nu[\lambda]}^{\nu[j]}[\vec{\theta}] $
        \If {$ \neg (c_{\sigma^*}^{j^*}[\vec{\theta}] + \vec{\theta^*} \leq \vec{\Theta}) $} \label{algo:sj:check_resources}
          \Continue
        \EndIf
        \State $ r \gets c_{\sigma^*[\lambda]}^{j^*}[\tau] \times  \rho^*$
        \If {$ r \geq \left| \Delta \right|$ }
          \State $\mu \gets \mu \cup \{ \nu \langle \sigma^*, \sigma^*[\lambda],j^* \rangle \} $ \label{algo:sj:full_segment_begin}
          \State $\rho^* \gets \rho^* - \frac{\left| \Delta \right|}{c_{\sigma^*[\lambda]}^{j^*}[\tau]}$ \label{algo:sj:full_segment_end} \label{algo:sj:rr_up_2}
        \Else
          \State $ \mu_1 \times \Delta_1, \mu_2 \times \Delta_2 \gets $ \Call{Split}{$\mu \times \Delta, \underline{\Delta} + r$ } \label{algo:sj:part_seg_begin} \label{algo:sj:split}
          \State $\mu_1 \gets \mu_1 \cup \{ \nu \langle \sigma^*, \sigma^*[\lambda],j^* \rangle \} $
          \State $\kappa \gets (\kappa \setminus \{\mu \times \Delta\}) \cup \{\mu_1 \times \Delta_1, \mu_2 \times \Delta_2 \} $
          \State $\rho^* \gets 0$,  \label{algo:sj:rr_up_1}
                 $t^f \gets \overline{\Delta_1}$ 
          \State \Break \label{algo:sj:part_seg_end}

        \EndIf
        \If { $\rho^* = 0 $ } 
          $t^f \gets \overline{\Delta}$, 
          \Break
        \EndIf
      \EndFor \label{algo:sj:segments_loop_end}

      \If {$\rho^* \neq 0$} \label{algo:sj:new_seg_begin}
        \State $ r \gets c_{\sigma^*[\lambda]}^{j^*}[\tau] \times  \rho^*$, 
               $\Delta \gets [t^e, t^e + r)$  
               \State $\mu \gets \{ \nu \langle \sigma^*, \sigma^*[\lambda],j^* \rangle \} $,
               $\kappa \gets \kappa \cup \{\mu \times \Delta\}$
        \State $t^e \gets t^e + r$,
               $\rho^* \gets 0$, 
               $t^f \gets \overline{\Delta}$ 
      \EndIf \label{algo:sj:new_seg_end}

      \If {$t^f > \sigma^*[\delta]$}
        \Return $\varnothing$ \label{algo:sj:deadline_check}
      \EndIf
      \State $\widetilde{\Sigma} \gets \widetilde{\Sigma} \setminus \sigma^*$
    \EndWhile \label{algo:sj:main_loop_end}
    \State \Return $\kappa$
  \end{algorithmic}
\end{algorithm}

Algorithm~\ref{algo:schedule_jc} takes the job configurations as input and generates a feasible schedule on.
Line~\ref{algo:sj:init} initialize the algorithm.
The algorithm iterates over unmapped jobs (lines~\ref{algo:sj:main_loop_begin}-\ref{algo:sj:main_loop_end}) in non-decreasing order of their deadlines, i.e., Earliest Deadline First (EDF) (line~\ref{algo:sj:next_job_edf}).
The loop in lines~\ref{algo:sj:segments_loop_begin}-\ref{algo:sj:segments_loop_end} schedules the job on already constructed mapping segments in the ascending order of the time segments (with a slight abuse of notation). 
After checking resource constraints on the segment (line~\ref{algo:sj:check_resources}),
the algorithm checks whether the job will execute during the whole segment (lines~\ref{algo:sj:full_segment_begin}-\ref{algo:sj:full_segment_end}) or only a part of it (lines~\ref{algo:sj:part_seg_begin}-\ref{algo:sj:part_seg_end}).
In the last case, the mapping segment is split at time the job finishes execution (line~\ref{algo:sj:split}), and the job is added only to the first part of it.
To track the remaining progress rate of the job while iterating the mapping segments, the algorithm initialize $\rho^*$ in line~\ref{algo:sj:rr_init} and  updates it in lines~\ref{algo:sj:rr_up_2} and \ref{algo:sj:rr_up_1}.
If the job is not finished after the last mapping segment, the new mapping segment is created and added to the schedule (lines~\ref{algo:sj:new_seg_begin}-\ref{algo:sj:new_seg_end}).
At the end, line~\ref{algo:sj:deadline_check} verifies that the job meets its deadline.
As a result, the algorithm puts more time-critical jobs into the earliest mapping segments as possible (EDF policy).

Note that the proposed algorithm is backward-compatible with the single-threaded version of the algorithm (without predictions)~\cite{niknafs19}, and the generated schedules will be the same due to MDF and EDF policies.
At the same time, due to a constraint to schedule all the threads on the same mapping segments, the original single-threaded algorithm cannot be employed for multi-threaded applications.

\section{Evaluation}
\label{sec:eval}
\begin{table}
\centering
\ra{1.1}
\caption{Amount of test cases differentiated by a number of jobs and deadline level.}
\label{tab:eval_test_distr}
\begin{tabular}{@{}l|cccc@{}}
\toprule
\diagbox[trim=l]{Deadline level}{\# Jobs}  & 1 & 2 & 3 & 4 \\
\midrule
Weak & 15 & 255& 255 & 230 \\
Tight & 35 & 340 & 340 & 206 \\
\bottomrule
\end{tabular}
\vspace{-3mm}
\end{table}

This section describes the experimental setup, the generation of the experimental workload and the alternative algorithms.
We evaluate the scheduling success rate, the energy-efficiency of found schedules, and the overhead of our approach. 

\subsection{Experimental setup and test generation}
\label{sec:eval_expsetup}

In our experiments we used three different dataflow applications from the automotive and multimedia domains: 
an algorithm of \textit{speaker recognition} with 8 processes~\cite{bouraoui2019comparing},
\textit{audio filter}, a stereo frequency filter with 8 processes~\cite{scopes17_tetris},
and an algorithm of \textit{pedestrian recognition} with 6 processes, provided by Silexica.
To obtain application configurations we exhaustively benchmarked these applications with input data of different sizes
on the Hardkernel Odroid XU4 featuring an Exynos 5422 big.LITTLE chip with four Cortex-A15 and four Cortex-A7 cores, fixed at frequencies of \unit[1.8]{GHz} and \unit[1.5]{GHz} respectively.
We measured the power consumption of the Odroid-XU4 board using ZES Zimmer LMG450 Power Analyzer connected to DC input with an external readout rate of \unit[20]{Sa/s}.
To identify Pareto-optimal configurations we executed the variants 50 times to get average execution times and energy consumptions. 
In this way we obtained 36 Pareto-configurations for audio filter, 35 for pedestrian recognition, and 28 for speaker recognition. 

The multi-application setup consists of 1676 test cases.
Each test has one to four jobs, which are characterized by the current progress ratio and the remaining deadline.
$31.9\%$ of the test cases consist of requests of a single application (uniformly distributed among each application and input data),
 while the remaining $68.1\%$ are application mixes.
In around $22.6\%$ of the tests we set the progress state of the jobs to zero (initial state).
For all others, we randomly choose a progress rate in the range $0-0.9$ except for the first job, which naturally starts in the initial state.
To set deadlines, we randomly select a configuration, calculate the remaining time to finish the job using this configuration and then scale it by a factor.
In the case of weak deadlines, we randomly choose large factors in the range $2-6$.
For tighter deadlines factors are selected in the range $0.6-2$ at random.
Table~\ref{tab:eval_test_distr} reports the amount of tests for each pair of number of jobs and deadline level.

Our proposed algorithm \textbf{MMKP-MDF} (Section~\ref{sec:algo}) was implemented in Python~3 and runs on a \unit[3.20]{GHz} Intel Core i5-6500 CPU.
Our so implemented RM prototype receives the Pareto-optimal configurations of the applications, reads a test case, and maps the applications on the Odroid XU4 platform.
We implemented alternative algorithms in the RM prototype to evaluate our solution. 
\textbf{EX-MEM} exhaustively checks all possible mappings for each of the mapping segments.
In each constructed mapping segment it cuts the segment on the shortest job, and generates the next mapping segment.
To accelerate the algorithm, we use memoization by storing and re-using the best energy consumption for a given current state (a pair of jobs, their progress rates, and time).
\textbf{MMKP-LR} is based on the Lagrangian Relaxation algorithm described in~\cite{wildermann15}.
This method solves Lagrangian relaxations of the MMKP problem using a subgradient method (limited by 100 iterations), and iteratively maps applications in the increasing order of the minimum configuration costs.
Similarly, during job mapping, the algorithm iteratively checks the configurations in the increasing order of their cost.
A configuration is mapped if there are enough resources and the job can meet the deadline either using this configuration till the end, or reconfigured to another configuration at the end of the mapping segment (optimistic check).
This process is repeated for the next mapping segment.
Thus, the analysis scope is limited to a single mapping segment.

\subsection{Scheduling rate and energy-efficiency}
\label{sec:eval_ee}

\begin{figure}[t]
  \centering
  \vspace{-3mm}
  \includegraphics[width=0.5\textwidth]{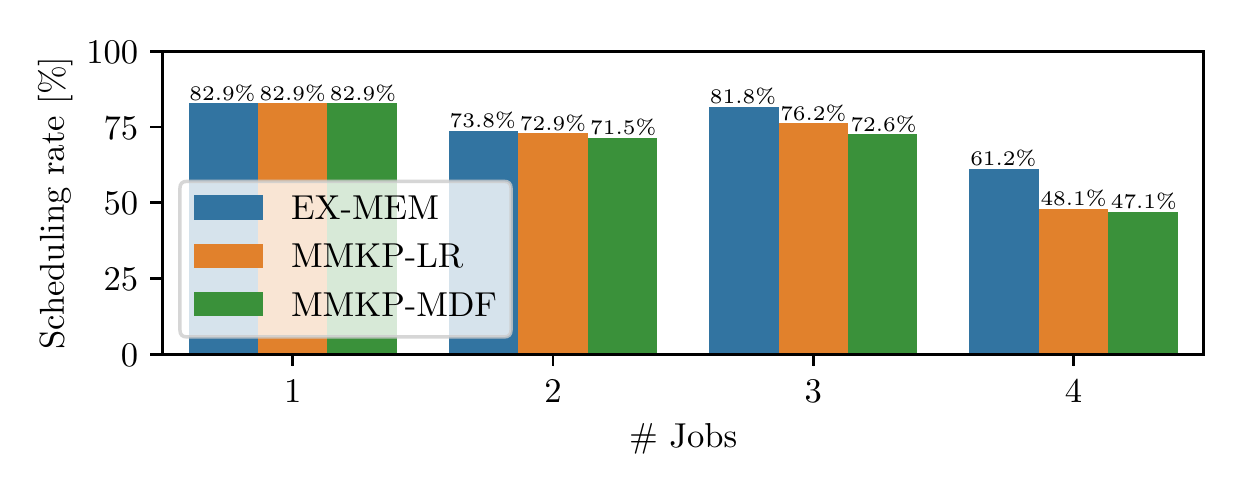}
  \vspace{-8mm}
  \caption{Scheduling rate of different schedulers for test cases with tight deadlines.}
  \label{fig:eval_acceptance}
  \vspace{-2mm}
\end{figure}

We evaluated all three implemented algorithms w.r.t the percentage of test cases they could find a feasible schedule.
All algorithms scheduled $100\%$ of the test cases with weak deadlines.
The results are completely different for the tests with tight deadline as shown in Fig.~\ref{fig:eval_acceptance}.
For test cases with one or two jobs, all three schedulers feature a similar scheduling success rate, with a difference within $2.3\%$.
For the tests with more jobs, EX-MEM shows significantly higher rate than the other two algorithms, up to $14.1\%$.
In all test cases, MMKP-LR and our MMKP-MDF achieve a similar scheduling success rate, with a difference within $3.6\%$ in favour of MMKP-LR.

\begin{table}
\centering
\ra{1.1}
\caption{Geometric mean of the relative energy consumption compared to EX-MEM.}
\label{tab:eval_gmean}
\begin{tabular}{@{}crrcrr@{}}
\toprule
&  \multicolumn{2}{c}{\textbf{MMKP-LR}} & \phantom{abc} & \multicolumn{2}{c}{\textbf{MMKP-MDF}} \\
\cmidrule{2-3} \cmidrule{5-6}
\# Jobs & Weak & Tight && Weak & Tight \\
\midrule
	1 & 1.0000 & 1.0000 && 1.0000 & 1.0000 \\
	2 & 1.0480 & 1.1291 && 1.0003 & 1.0682 \\
	3 & 1.1534 & 1.2250 && 1.0031 & 1.0978 \\
	4 & 1.2648 & 1.3404 && 1.0099 & 1.0618 \\
\midrule
\textbf{Overall} & \textbf{1.1452} & \textbf{1.1923} && \textbf{1.0042} & \textbf{1.0756} \\
  \textbf{(all levels)} & \multicolumn{2}{c}{\textbf{1.1665}} && \multicolumn{2}{c}{\textbf{1.0356}} \\
\bottomrule
\end{tabular}
\end{table}

\begin{figure}[t]
  \centering
  \vspace{-3mm}
  \includegraphics[clip,width=\columnwidth]{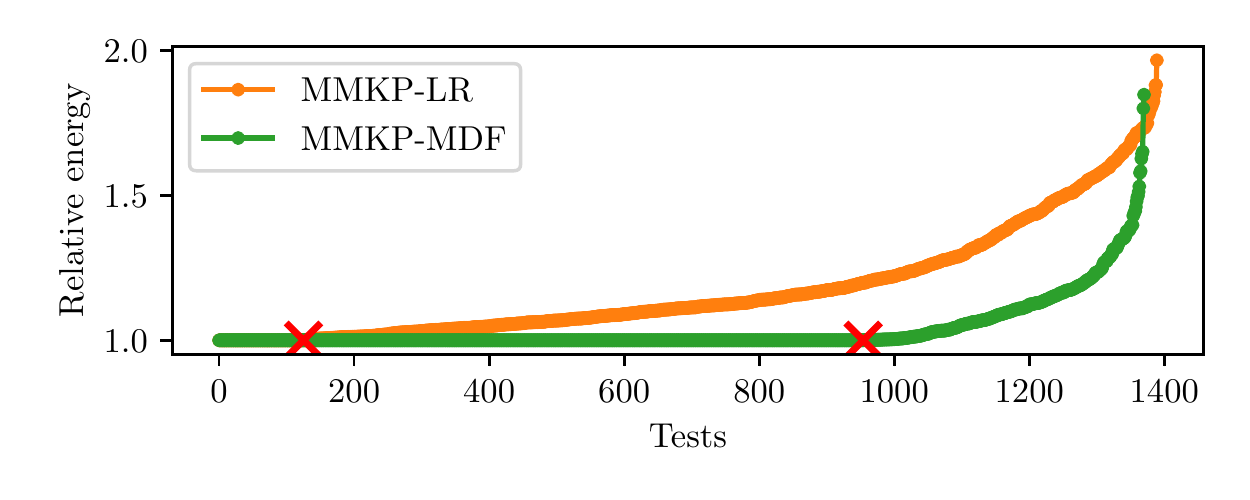}%
  \vspace{-3mm}
  \caption{S-curves of the relative energy consumption compared to EX-MEM (lower is better).}
  \label{fig:eval_scurve}
  \vspace{-2mm}
\end{figure}

In terms of energy efficiency, we compared the algorithms to the optimal solutions obtained by EX-MEM. 
For each successfully scheduled test case, we compute the relative energy consumption compared to EX-MEM, and report the geometric mean of these values for each test group in Table~\ref{tab:eval_gmean}.
All schedulers generate optimal schedules in case of a single job.
For tests with weak deadlines, the relative energy consumption of MMKP-MDF schedules increases slowly from $0.03\%$ for two jobs till $0.99\%$ for four jobs (in geometric mean), and over all tests with weak deadline, the schedules are off by $0.42\%$ from the optimal one. 
For tighter deadlines, the relative energy consumption of MMKP-MDF varies nonmonotonically with the number of jobs, and they are off by $7.56\%$ in geometric mean.
For MMKP-LR, the relative energy consumption increases with the number of jobs, and overall the geometric mean of these values are $14.52\%$ and $19.23\%$ for weak and tight deadlines, correspondingly.
Overall, MMKP-MDF generates more energy-efficient schedules by $13.1\%$ than MMKP-LR.
Fig.~\ref{fig:eval_scurve} presents S-curves of relative energy consumption over all tests.
As we see, MMKP-MDF generates optimal schedules for $954$ tests ($69.6\%$ of successfully scheduled), while MMKP-LR only for $125$ tests ($9.0\%$).

\subsection{Search time}
\label{sec:eval_time}

\begin{figure}[t]
  \centering
  \includegraphics[clip,width=\columnwidth]{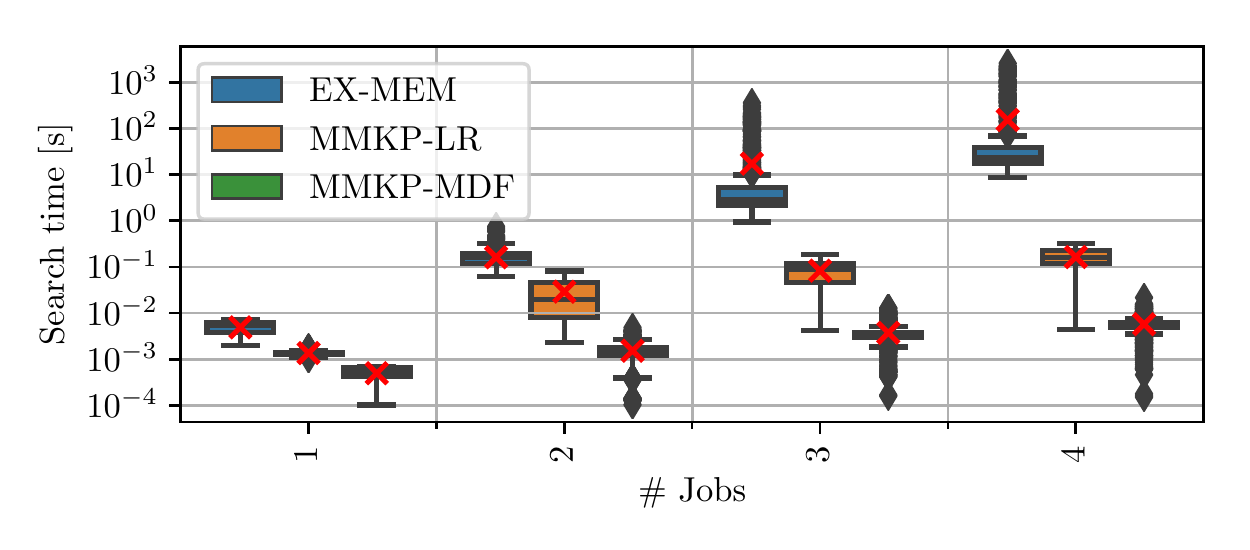}%
  \vspace{-3mm}
  \caption{Box plots and the average values summarizing the scheduling overhead for different algorithms.}
  \label{fig:eval_search}
  \vspace{-2mm}
\end{figure}

Fig.~\ref{fig:eval_search} shows the boxplots and the average values of the execution times of different algorithms differentiated by the number of jobs.
As we see in the figure, the scheduling time increases with the number of jobs for all three implementations.
As expected, EX-MEM displays an exponential growth, with an average of \unit[152]{s} to schedule four jobs, while the median and the worst-case values are \unit[22.65]{s} and \unit[2550]{s} ($\approx \unit[37.5]{min}$) correspondingly.
The scheduling time of MMKP-LR and MMKP-MDF grows less rapidly.
MMKP-LR needs around \unit[1.3]{ms} to schedule one job and around \unit[163]{ms} for four jobs.
MMKP-MDF is significantly faster, requiring only \unit[5.7]{ms} in average for four jobs with a worst-case of \unit[21.6]{ms}.

To summarize, our proposed MMKP-MDF algorithm achieves comparable scheduling success rate as the MMKP-LR scheduler, while outperforming it in terms of overall energy efficiency and scheduling overhead. 
MMKP-MDF schedules applications within \unit[21.6]{ms}, which makes it a good candidate to implement in a fully-functional resource runtime manager.
As mentioned in the experimental setup, the overhead analysis for all schedulers was performed on a prototyped RM written in Python~3. 
Better performances can be expected from a C implementation.

\section{Conclusion}
\label{sec:conc}
We investigated how mapping analysis with global scope can improve the quality of generated schedules for multi-threaded firm real-time applications.
We proposed a fast algorithm to schedule the applications on  heterogeneous multi-core systems.
The proposed approach achieves a scheduling rate competitive with the state-of-the-art, while improving the energy efficiency by around $13\%$.
The generated solutions are only $3.6\%$ off from optimal schedules obtained exhaustively.
The algorithm runs an order of magnitude faster than the state-of-the-art approach, making it a good candidate for integration into other runtime resource managers.

\section*{Acknowledgment}
This work was supported in part by the German Research Foundation (DFG) within the Collaborative Research Center HAEC and the Center for Advancing Electronics Dresden (cfaed).
We thank Silexica (\url{www.silexica.com}) for making their SLX Tool Suite and the applications available to us.

\IEEEtriggeratref{2}
\bibliographystyle{IEEEtran}
\bibliography{IEEEabrv,paper}

\end{document}